\documentclass[12pt,twoside]{article}
\begin{document}
\baselineskip=0.20in
\vspace{20mm}
\baselineskip=0.30in
\begin{center}

{\large \bf Solutions of the Dirac equation with spin and pseudospin Symmetry for the trigonometric Scarf potential in $D$-dimensions.}

\vspace{4mm}

{ \large {\bf Falaye, B.\ J.}\footnote{E-mail:~ fbjames11@physicist.net}  and {\bf Oyewumi, K.\,J.}\footnote{E-Mail:~~ kjoyewumi66@unilorin.edu.ng}} \vspace{5mm}

{Theoretical Physics Section, Department of Physics\\ University of Ilorin,  P. M. B. 1515, Ilorin, Nigeria. }

\vspace{4mm}

\end{center}

\noindent
\begin{abstract}
\noindent
Solutions of the Dirac equation with spin and pseudospin symmetry for the scalar and vector trigonometric scarf potential in $D$-dimensions within the framework of an approximation scheme to the centrifugal barrier are obtained. The energy spectral and the two-component spinor eigenfunctions are obtained. The bound state energy levels for various values of dimension $D$, $n$, $\kappa$ and the potential range parameter $\alpha$ are also presented.
\end{abstract}

{\bf Keywords}: Dirac equation; Nikiforov-Uvarov; Scarf; spin symmetry; pseudospin symmetry; eigenvalues; spinors

{\bf PACs No.} 03.65.Ge; 03.65.Pm

\section{Introduction}
In other to investigate nuclear shell structure, the study of spin and pseudospin symmetric solutions of the Dirac equation has been an important area of research in nuclear physics. The concept of spin and pseudospin symmetry with nuclear shell model has been used widely in explaining a number of phenomena in nuclear physics and related area.

Within the framework of the Dirac equation, the spin symmetry occurs when the difference of the potential between the repulsive Lorentz vector potential V(r) and attractive Lorentz scalar potential S(r) is a constant, that is, $\Delta(r) = V(r) - S(r) = \mbox{constant}$. However, pseudospin symmetry arises when sum of the potential of the repulsive Lorentz vector potential V(r) and attractive Lorentz scalar potential S(r) is a constant, that is, $\Sigma(r) = V(r) + S(r) = \mbox{constant}$ \cite{ArE69, HeA69, Gin97, LeG01, Gin04, Gin051,Gin052}. 

The spin symmetry is relevant for meson \cite{PaE01}. The pseudospin symmetry has been used: to study the structure of the deformed nuclei \cite{DuE87, BoE82}, to construct an effective shell model coupling scheme \cite{TrE95} and identical bands and triaxality  observed in nuclei \cite{BeE97}.

Recently many researchers have applied spin and pseudospin symmetry conditions on numbers of potentials. These potentials include: Hulth$\grave{e}$n potential \cite{Sad07, SoE07}, Eckart potential \cite{JiE06, ZhE08, SoE081, SoE082}, P$\ddot{o}$schl-teller potential  \cite{JiE07, Agb11}, the Rosen- Morse potential \cite{OyA10,Ikh10}, three parameter diatomic potential \cite{JiE07}, harmonic potential \cite{GuE05}, Pseudoharmonic potential \cite{AyS09}, Manning-Rosen potential \cite{ChE09}, Wood-Saxon potential \cite{GuS05} and kratzer potential with angle dependent potential \cite{BeS09}.

On the other hand, exact analytical solutions for $\ell \neq 0$ states are not possible for some of these potentials due to the centrifugal term. An approximation to the (pseudo or) - centrifugal term has been used by many researchers to obtain the approximate analytical solutions \cite{SoE07, SoE081, SoE082, JiE09, Ikh10, WeD10s}. Several researchers have used various methods to solve spin and pseudopsin symmetry problems including the centrifugal approximation ranging from the Asymptotic Iteration method (AIM), the Nikiforov-Uvarov method (N-U), Supersymmetric and shape invariance method to functional analysis method. Due to its wider applications and simplicity , we are adopting the Nikiforov-Uvarov method \cite{NiU88, NiE91}. This method find its usefullness in the study of relativistic and non-relativistic quantum mechanics, see Ikhdair et al. (2011) and the references therein \cite{IkE11}. 

In this study, we are considering the Scarf potential \cite{Yes07}
\begin{equation}
V(r)=  - \frac{V_{0}}{\sin^{2}{\alpha r}} 
\label{s1}
\end{equation}
where $V_{0}$ is the depth of the Scarf potential and $\alpha$ is the range of the potential.

Futhermore, with interest in higher dimensional spaces, the multidimensional relativistic and non-relativistic equations have been investigated by many authors. This investigation have employed Isotropic harmonic oscillator plus inverse quadratic potential \cite{OyB03}, Pseudoharmonic potential \cite{OyE08}, Kratzer-Fues potential \cite{Oye05,Agb112} and hydrogen atom \cite{Alj98} (for comprehensive reviews on abtrary dimension, see Dong 2007).
In the present work, Section $2$ contained a brief review of the Nikiforov-Uvarov (N-U) method. The Dirac equation in $D$-dimension is described in Section $3$. Section $4$ contains the bound state solutions of the Dirac equation with the trigonometric Scarf potential and conclusion is contained in Section $5$.

\section{The Nikiforov-Uvarov (N-U) Method}
The N-U method is based on solving a second order linear differential equation by reducing it to a generalized equation of hypergeometric type \cite{NiU88, NiE91}. By introducing an appropriate coordinate transformation $z = z(r)$, this equation can be re-written in the following form.
\begin{equation}
\Psi ''(z) + \frac{\tilde{\tau}(z)}{\sigma(z)} \Psi '(z) + \frac{\tilde{\sigma}(z)}{\sigma^{2}(z)}\Psi(z)=0
\label{s2} 
\end{equation}	
where $\sigma(z)$ and $\tilde{\sigma}(z)$ are polynomials, at most of second degree, and $\tilde{\tau}(z)$  is a first degree polynomial. Now, if one takes the following factorization	
\begin{equation}
\Psi(z) = y(z) \phi(z) .
\label{s3}
\end{equation}	

The equation (\ref{s3}) reduces to a hypergeometric type equation
\begin{equation}
\sigma(z) y''(z) + \tau(z)y'(z) + \lambda y(z) = 0 
\label{s4},
\end{equation}
where
\begin{equation}
\tau(z) = \tilde{\tau}(z) + 2 \pi(z) 
\label{s5}.
\end{equation}
The function $\pi$ and the parameter $\lambda$ require for this method are define as follows
\begin{equation}
\pi(z) = \left(\frac{\sigma'(z) - \tilde{\tau}(z)}{2}\right) \pm  \sqrt{\left( \frac{\sigma'(z) - \tilde{\tau}(z)}{2}\right)^{2} + \tilde{\sigma}(z) + k \sigma(z) }
\label{s6}
\end{equation}
and
\begin{equation}
\lambda = k + \pi'(z)
\label{s7} .
\end{equation}

In other to find the value of $k$, the expression under the square root must be a square of a polynomial. This gives the polynomial $\pi(z)$  which is dependent on the transformation function $z(r)$. Also the parameter $\lambda$ defined in equation (\ref{s7}) takes the form
\begin{equation}
\lambda =\lambda_{n_r} = - n_{r} \tau' - \left[ \frac{n_{r} (n_{r} - 1)}{2} \sigma '' \right]
\label{s8}.
\end{equation}
The polynomial solutions $y_{n}(z)$ are given by the Rodrigue relation
\begin{equation}
y_{n_{r}}(z) = \frac{N_{n_{r}}} {\rho(z)} \frac{d^{n_{r}}}{dz^{n_{r}}} \left[\sigma^{n_{r}}(z) \rho(z) \right] 
\label{s9},
\end{equation}
where $N_{n_{r}}$ is the normalization constant and $\rho(z)$ is the weight function satisfying
\begin{equation}
\frac{d}{dz}\left[\sigma(z) \rho(z) \right] = \tau(z)\rho(z)
\label{s10}.
\end{equation}
Second part of the wave function $\phi(z)$  can be obtained from the relation
\begin{equation}
\pi(z) = \sigma(z) \frac{d}{dz}[\ln \phi(z)] 
\label{s11}.
\end{equation}

\section{ The Dirac Equation in $D$- Dimensions}
The $D$-dimension Dirac equation for a particle of mass $M$, with radially symmetric Lorentz vector $V(r)$ and Lorentz scalar potential $S(r)$ is given(in atomic units $\hbar = c = 1$ \cite{ Agb11, BjD64, Gre00, AlE06, CiE05}:
\begin{equation}
i\frac{ \partial \Psi(\vec{r})}{dt} = H \Psi(\vec{r}), H = \sum_{j = 1}^{D} \alpha _{j}P_{j} +  \hat{\beta } \left[M + S(r) \right] + V(r),
\label{s12} 
\end{equation}
where $M$ is the mass of the particle, $V(r)$ and $S(r)$ are spherically symmetric vector and scalar potentials, respectively, $P$ is the momentum operator, $\{\alpha_{i}\} $ and $\hat{\beta}$ are Dirac matrices satisfying anti-commutation relations. 

Following the procedure stated by Ciftci et al. (2005) and Agboola (2011), the spinor wave functions can be written using the Pauli-Dirac representation as:
\begin{equation}
\Psi_{n_{r},\kappa}(\vec{r}, \Omega_{D})=  
\frac{1}{r^{\frac{D - 1}{2}}}\left[\matrix {
F_{n\kappa}(r) ~Y^{\ell}_{jm}(\Omega_{D})\cr
iG_{n\kappa}(r)~Y^{\overline{\ell}}_{jm}(\Omega_{D})\cr
}\right],~ \kappa = \pm\left(j + \frac{D - 2}{2}\right) 
\label{s13}
\end{equation}
where $F_{n_{r}\kappa}(r)$ and $G_{n_{r}\kappa}(r)$ are the radial wave functions of the upper- and the lower -spinor components, respectively. $Y_{jm}^l (\Omega_D)$ and $Y_{jm}^{\tilde{l}}(\Omega_D)$ are the hyperspherical harmonic functions coupled with the total angular momentum, while, $\ell$ and $\tilde{l}$ correspond to the orbital and pseudo-orbital angular momentum quantum numbers for spin and pseudospin symmetry conditions. $\kappa = -\left( j + \frac{D -2}{2}\right)$ for aligned spin $j = \ell + \frac{1}{2}$ and $\kappa = \left(j + \frac{D - 2}{2}\right)$ for unaligned spin $j = \ell - \frac{1}{2}$.

Substituting Equation(\ref{s13}) into Equation (\ref{s12}), the two coupled second-order differential equation for the upper and lower component spinors are obtained follows:
\begin{equation}
\left( \frac{d}{dr} + \frac{\kappa}{r} \right)F_{n_{r},\kappa}(r) = \left[M + E_{n_{r},\kappa} + S(r) - V(r) \right]G_{n_{r},\kappa}(r) 
\label{s14},
\end{equation}
\begin{equation}
\left( \frac{d}{dr} + \frac{\kappa}{r} \right)G_{n_{r},\kappa}(r) = \left[M - E_{n_{r},\kappa} + S(r) + V(r) \right]F_{n_{r},\kappa}(r) 
\label{s15}~.
\end{equation}
Substituting the expression of $G_{n_{r}, \kappa}(r)$ obtained from equation (\ref{s15}) into equation (\ref{s14}), we have two second-order differential equations for the upper and lower component spinors as: 
\begin{eqnarray}
&\displaystyle{\left\{- \frac{d^{2}}{dr^{2}} + \frac{\kappa(\kappa + 1)}{r^{2}} + \left[M + E_{n_{r}, \kappa} - \Delta(r)\right] \left[M - E_{n_{r}, \kappa} + \Sigma(r)\right] \right\}F_{n_{r}, \kappa}(r)} \nonumber \\
& \displaystyle{ = \frac{\frac{d \Delta(r)}{dr} (\frac{d}{dr} + \frac{\kappa}{r} )}{\left[M + E_{n_{r}, \kappa} - \Delta(r)\right]} F_{n_{r}, \kappa}(r)}
\label{s16},
\end{eqnarray}
\begin{eqnarray}
&\displaystyle{\left\{- \frac{d^{2}}{dr^{2}} + \frac{ \kappa(\kappa - 1)}{r^{2}} + \left[M + E_{n_{r}, \kappa} - \Delta(r)\right] \left[M - E_{n_{r}, \kappa} + \Sigma(r)\right] \right \}G_{n_{r}, \kappa}(r)} \nonumber \\
&\displaystyle{ = -\frac{\frac{d \Sigma(r)}{dr} (\frac{d}{dr} - \frac{\kappa}{r} )}{\left[M - E_{n_{r}, \kappa} + \Sigma(r)\right]}G_{n_{r}, \kappa}(r)} 
\label{s17},
\end{eqnarray}
where $\Delta(r) = V(r) - S(r)$ and $\Sigma(r) = V(r) + S(r)$ are the difference and the sum of the potentials $V(r)$ and $S(r)$, respectively and $\kappa = \pm \left(\frac{2\ell + D - 1}{2}\right)$ \cite{ Agb11, BjD64, CiE05}. 

In the  presence of the spin symmetry condition, that is, the difference potential $\Delta(r)= V(r) - S(r) = C_{s} = \mbox{constant or } \frac{d\Delta (r)}{dr} = 0$, then, equation (\ref{s16}) reduces into
\begin{eqnarray}
&\displaystyle{\left\{- \frac{d^{2}}{dr^{2}} + \frac{\kappa(\kappa + 1)}{r^{2}} +  \left[ M + E_{n_{r}, \kappa} - C_{s} \right] \Sigma(r)  \right \} F_{n_{r}, \kappa}(r)} \nonumber \\
& = \displaystyle{\left[E_{n_{r}, \kappa}^{2} - M^{2} + C_{s} (M - E_{n_{r}, \kappa}) \right]F_{n_{r}, \kappa}(r)}
\label{s18}.
\end{eqnarray}
Then, the lower component $G_{n_{r},\kappa}(r)$ of the Dirac spinor is obtained as
\begin{equation}
G_{n_{r}, \kappa} (r)= \frac{1}{M + E_{n_{r}, \kappa} - C_{s}}\left[\frac{d}{dr} + \frac{\kappa}{r} \right]F_{n_{r},\kappa} (r)
\label{s19},
\end{equation}
where $E_{n_{r},\kappa} + M \neq 0$, only real positive energy state exist when $C_{s} = 0$ \cite{Ikh10, GuS05, IkE11}.

Again, under the pseudospin symmetry condition, that is, the sum potential $\Sigma(r) = V(r) + S(r) = C_{ps}$ $ \mbox{constant or}~ \frac{d\Sigma(r)}{dr} = 0$, then, equation (\ref{s17}) can be written as
\begin{eqnarray}
&\displaystyle{\left\{- \frac{d^{2}}{dr^{2}} + \frac{\kappa(\kappa - 1)}{r^{2}} - \left[ M - E_{n_{r}, \kappa} + C_{ps} \right] \Delta(r)  \right \} G_{n_{r}, \kappa}(r)} \nonumber \\
&\displaystyle{ =\left[E_{n_{r}, \kappa}^{2} - M^{2} + C_{ps} (M - E_{n_{r}, \kappa}) \right]G_{n_{r}, \kappa}(r)}
\label{s20},
\end{eqnarray}
where the upper component $F_{n_{r}, \kappa}(r)$ is obtained as
\begin{equation}
F_{n, \kappa} (r)= \frac{1}{Mc^{2} - E_{n\kappa} + C_{ps}}\left[\frac{d}{dr} - \frac{\kappa}{r} \right]G_{n\kappa} (r)
\label{s21},
\end{equation}
where $E_{n\kappa} - Mc^{2} \neq 0$, only real negative energy state exist when $C_{ps} = 0$ \cite{Ikh10, GuS05, IkE11}.

These equations cannot be solved exactly due to centrifugal term, so, we used the following approximation
\begin{equation}
\frac{1}{r^{2}} \approx \frac{\alpha^{2}}{\cos^{2} \alpha r} 
\label{s22}.
\end{equation}
With this approximation, the second order differential equations in equation (\ref{s18}) and (\ref{s20}) can be solved approximately by using Nikiforov-Uvarov method.

\section{Bound State Solutions of the Trigonometric Scarf potential}

\subsection{Spin symmetry solutions of the Dirac equation with the Trigonometry Scarf potential with arbitrary $\kappa$ in D dimension}

Under the condition of the spin symmetry,  $\Delta(r) = V(r) - S(r) = C_{s}$, the vector and scalar trigonometric Scarf potential is considered \cite{Yes07}. Then,  
\begin{equation}
V(r) =  - \frac{V_{o}}{\sin^{2} \alpha r}~ \mbox{and} ~S(r) =  - \frac{S_{o}}{\sin^{2}\alpha r} 
\label{s23}
\end{equation}
where $V_{o}$  and $S_{o}$ are the depth of vector and scalar trigonometric Scarf potential, therefore,
\begin{equation}
\sum(r) = - \frac{(V_{o} + S_{o})}{\sin^{2} \alpha r} 
\label{s24}.
\end{equation}

Substituting equations (\ref{s22}) and (\ref{s24}) into equation (\ref{s18}), we have
\begin{equation}
\left[ \frac{d^{2}}{d r^{2}} - \frac{ \gamma \alpha^{2}}{r^{2}} - \varepsilon^{2} \alpha^{2} + \frac{ \eta \alpha ^{2}}{ \sin^{2} \alpha r} \right] F_{n_{r},k}(r) = 0,
\label{s25}
\end{equation}
where 
\begin{eqnarray}
&\gamma = k(k + 1), \nonumber \\
&\varepsilon^{2}\alpha^{2} = \left(M - E_{n_{r}, k} \right) \left(M + E_{n_{r}, k} - C_{s} \right), \nonumber \\
&\eta \alpha^{2} = (V_{o} + S_{o}) \left(M + E_{n_{r}, k} - C_{s} \right). 
\label{s26}
\end{eqnarray}

In other to apply N-U method, we introduced a new variable
\begin{equation}
z = - \tan^{2} {\alpha r} 
\label{s27}
\end{equation}
With equation (\ref{s27}), equation (\ref{s25}) becomes
\begin{equation}
F''_{n_{r}, k}(z) + \frac{1 - 3 z}{2z(1 - z)} F'_{n_{r}, k}(z) + \frac{1}{4 z^{2}(1 - z)^{2}}\left[- \gamma z^{2} + (\gamma  + \varepsilon^{2} - \eta)z + \eta \right]F_{n_{r}, k}(z)
\label{s28}.
\end{equation}

Substituting for $\sigma(z)$,$\tilde{\sigma}(z)$ and $\tilde{\tau}(z)$ in equation (\ref{s6}), we obtained function $\pi(z)$ as
\begin{equation}
 \pi(z) = \sqrt{z^{2} \left[( 1 + 4 \eta) + 4i \varepsilon \sqrt{(1 + 4 \eta )} - 4 \varepsilon^{2} \right] - z\left[ 2 + 8 \eta + 4i \varepsilon \sqrt{(1 + 4 \eta )} \right] + \left(1 + 4 \eta  \right)}
\label{s29}.
\end{equation}
According to the Nikiforov-Uvarov method, the expression under the square root must be a square of a polynomial and therefore, the new $\pi(z)$ function for each $k$ is obtained as
\begin{eqnarray}
 \pi(z) = 
 \left\{
\begin{array}{ll}
&\frac{z}{2}\left[\Lambda + (2i \varepsilon - 1) \right] + \frac{1}{2}\left[1 -  \Lambda \right] 
; k = - \frac{1}{2}\left[ \eta - \gamma - \varepsilon^{2}\right] - \frac{\Lambda i \varepsilon}{2}   \\
& -\frac{z}{2}\left[\Lambda + (2i \varepsilon - 1) \right] + \frac{1}{2}\left[1 +  \Lambda \right] 
; k = - \frac{1}{2}\left[ \eta - \gamma - \varepsilon^{2}\right] - \frac{\Lambda i \varepsilon}{2}   \\
&\frac{z}{2}\left[\Lambda - (2i \varepsilon + 1) \right] + \frac{1}{2}\left[1 -  \Lambda \right] 
; k = - \frac{1}{2}\left[ \eta - \gamma - \varepsilon^{2}\right] + \frac{\Lambda i \varepsilon}{2}   \\
&- \frac{z}{2}\left[\Lambda - (2i \varepsilon - 1) \right] + \frac{1}{2}\left[1 +  \Lambda \right] 
; k = - \frac{1}{2}\left[ \eta - \gamma - \varepsilon^{2}\right] + \frac{\Lambda i \varepsilon}{2}  
\end{array}\right.
\label{d30}
\end{eqnarray}
where $\Lambda = \sqrt{1 + 4\eta} $.

For the polynomial $\tau(z)=\tilde{\tau}(z)+2\pi$ , with a negative derivative, we select
\begin{equation}
\pi(z) = -\frac{z}{2}[\Lambda + (2 i \varepsilon - 1) ] + \frac{1}{2}[1 + \Lambda ];~~ k = - \frac{1}{2}[\eta - \gamma - \varepsilon^{2}] - \frac{\Lambda i \varepsilon}{2} 
\label{s31}
\end{equation}
and when this is combined with $\lambda = k + \pi'(z)$, gives
\begin{equation}
\lambda = - \frac{1}{2}[\eta - \gamma - \varepsilon^{2}] - \frac{\Lambda i \varepsilon}{2} - \frac{1}{2}[\Lambda + (2 i \varepsilon - 1) ] 
\label{s32}.
\end{equation}
Now using equation (\ref{s8}), we obtained 
\begin{equation}
\lambda = -2n_{r} + 2 i \varepsilon n_{r} + \Lambda n_{r} + 2n_{r}(n_{r} -  1),
\label{s33}
\end{equation}
and on comparing equations (\ref{s33}) and (\ref{s32}), we obtained the energy equation as
\begin{equation}
\varepsilon = \frac{i}{2}\left[ 2 (2n_{r} + 1) + (2k + 1) + \Lambda \right]
\label{s34}.
\end{equation}
Or more explicitly
\begin{equation}
(M - E_{n_{r}, \kappa})(M + E_{n_{r}, \kappa} - C_{s}) = -\frac{\alpha^{2}}{4}\left[2 n + D + \frac{1}{\alpha}\sqrt{\alpha^{2} + (4V_{o} +  S_{o})(M + E_{n_{r}} - C_{s})} \right]^{2}
\label{s35},
\end{equation}
where we have defined a principal quantum number as $n - \ell = 2n_{r} + 1$, chosen $k = l + \frac{(D-1)}{2}$  and $n = 1,2, ~~\ldots $. The explicit values of the energy for different values of $n_{r}$, $\alpha$ and $D$ are shown for spin symmetry in the Table 1. below:

\begin{table}[!hbp]
\caption{Bound-state energy level $E_{n_{r}}$ for different values of $n_{r}$, $\alpha$ and $D$ with $M = C_{s} = V_{o} = S_{o} = 1$.  \vspace*{13pt}} {\small
\begin{tabular}{|c|c|c|c|c|c|c|}\hline
{}&{}&{} &{} &{}&{} &{}\\[-1.5ex]
$D$&$n_{r}$&$\alpha =  0.0001$ & $\alpha = 0.0005$ & $\alpha = 0.001$ & $\alpha = 0.005$&$\alpha = 0.01$\\[2ex]\hline
& 1 &3.0004   &3.0020   &3.0041   &3.0204    &3.0408 \\[1ex]   
& 2 &3.0006   &3.0029   &3.0057   &3.0285    &3.0570 \\[1ex]  
3   & 3 &3.0007   &3.0037   &3.0073   &3.0367    &3.0733 \\[1ex]  
    & 4 &3.0009   &3.0045   &3.0090   &3.0448    &3.0895 \\[1ex] 
    & 5 &3.0011   &3.0053   &3.0106   &3.0350    &3.1057  \\[1ex] \hline
    &1  &3.0005   &3.0024	 &3.0049	 &3.0245   &3.0489\\[1ex]
    &2  &3.0007   &3.0033   &3.0065   &3.0326   &3.0652\\[1ex]
4   &3	 &3.0008   &3.0041	 &3.0082   &3.0408	 &3.0814\\[1ex]
    &4  &3.0010   &3.0049   &3.0092   &3.0489	 &3.0976\\[1ex]
    &5  &3.0011   &3.0057   &3.0114   &3.0570   &3.1138\\[1ex] \hline
    &1  &3.0006   &3.0029   &3.0057   &3.0285   &3.0570\\[1ex]
    &2  &3.0007   &3.0037   &3.0073   &3.0367   &3.0733\\[1ex]
5   &3  &3.0009   &3.0045   &3.0090   &3.0448   &3.0895\\[1ex]
    &4  &3.0011   &3.0053   &3.0106   &3.0530   &3.1057\\[1ex]
    &5  &3.0012   &3.0061   &3.0122   &3.0611   &3.1219\\[1ex] \hline
    &1  &3.0007   &3.0033	 &3.0065   &3.0326   &3.0652\\[1ex]
    &2  &3.0008   &3.0041	 &3.0082   &3.0408   &3.0814\\[1ex]
6   &3  &3.0010   &3.0049   &3.0098   &3.0489   &3.0976\\[1ex]
    &4  &3.0011   &3.0057   &3.0114   &3.0570   &3.1138\\[1ex]
    &5  &3.0013	 &3.0065   &3.0131   &3.0651   &3.1300\\[1ex] \hline
\end{tabular}\label{tab1} }
\vspace*{-13pt}
\end{table}

From the Nikiforov-Uvarov method, the hypergeometric function $y(z)$ depend on the determination of the weight function $\rho(z)$ which is satisfied by the differential equation $\frac{d}{dz}[\sigma(z)\rho(z)] = \tau(z) \rho(z)$. Thus $\rho(z)$ is obtained as
\begin{equation}
\rho(z) = (1-z)^{i \varepsilon}z^{\frac{\Lambda}{2}}, 
\label{s36}
\end{equation}
by substituting $\rho(z)$ into the Rodrigue relation (\ref{s9}) one gets
\begin{equation}
y_{n_{r}} = (1 - z)^{-i\varepsilon} z^{-\frac{\Lambda}{2}} \frac{d^{n_{r}}}{dz^{n_{r}}}\left[2^{n_{r}}(1 - z)^{n_{r} + i\varepsilon} z^{n_{r} + \frac{\Lambda}{2}} \right]
\label{s37}. 
\end{equation}
The solution of equation (\ref{s37}) may be expressed in terms of the Jacobi polynomials of the form \cite{GrR07}
\begin{equation}
y_{n_{r}} = N_{n_{r}} P_{n_{r}}^{(i \varepsilon,\frac{\Lambda}{2})}(1 - 2z) 
\label{s38}
\end{equation}
and $\phi(z)$ is obtained as
\begin{equation}
\phi(z) = (1 - z)^{i \varepsilon} z^{\frac{1 + \Lambda}{4}},
\label{s39} 
\end{equation}
therefore, the radial wavefunction of the upper component of the Dirac wavefunction $F_{n_{r}, k}(z)$ can be written as
\begin{equation}
 F_{n_{r},k}(z) = N_{n_{r}}(1 - z)^{i \varepsilon}z^{\frac{1 + \Lambda}{4}}P_{n_{r}}^{(i \varepsilon, \frac{\Lambda}{2} )}(1 - 2z).
 \label{s40} 
\end{equation}

The corresponding lower -component spinor $G_{n_{r}, k}(z)$ in equation (\ref{s19}) can be obtained by using the following relation \cite{AbS70}:
\begin{equation}
\frac{d}{dz}[P_{n}^{( a, b)}(z)] = \frac{1}{2}(a + b + n + 1)P_{n - 1}^{(a + 1, b + 1)}(z).
\label{s41} 
\end{equation}
Using above equation (\ref{s41}), we obtained the lower component spinor $G_{n_{r}, k}(z)$ from equation (\ref{s19}) as
\begin{eqnarray}
G_{n_{r}, k} (z) =      \left\{
\begin{array}{ll}
& \displaystyle{N_{n_{r}}(1 - z)^{\frac{i \varepsilon}{2}}z^{\frac{1 + \Lambda}{4}}P_{n_{r}}^{(i \varepsilon, \frac{\Lambda}{2} )}( 1 - 2z)\left[ - \alpha \left\{\left(i \frac{\Lambda + 1}{2} \right)\left( \frac{1 - z}{\sqrt{z}} \right) + \varepsilon \sqrt{z} - \frac{k}{i \tanh{\alpha }} \right\} \right]} \\
& \displaystyle{+ \ N_{n_{r}}(1 - z)^{\frac{i \varepsilon + 2}{2}} z^{\frac{3 + \Lambda}{4}}P_{n_{r} - 1}^{\left(i \varepsilon + 1, \frac{\Lambda}{2} + 1 \right)}(1 - 2z) \left[- \alpha \left\{ \left(i \frac{\Lambda + 2n_{r}}{2} \right) + \varepsilon  \right\} \right] }~~,
\end{array}\right.  
\label{d37}
\end{eqnarray}

where $N_{n_{r}}$ is the normalization factor to be determined from normalization condition
\begin{equation}
\int_{0}^{\infty} \mid F_{n_{r}, k}(r)\mid^{2}dr = - \frac{1}{2 i\alpha} \int_{0}^{1}F_{n_{r}, k} (z)[ \sqrt{z}(1-z)]^{2}dz = 1. 
\label{s43}
\end{equation}

This can further be written as
\begin{equation}
 1 = \frac{N_{n_{r}}^{2}}{- 2i \alpha } \int_{0}^{1}z^{\frac{\Lambda}{2}}(1-z)^{i \varepsilon -1}\left[ P_{n_{r}}^{( i \varepsilon, \frac{\Lambda}{2})}(1 - 2z) \right]^{2}dz.
 \label{s44} 
\end{equation}
Two different forms of the Jacobi polynomials are \cite{AbS70, GrR07, MaE66}:
\begin{equation}
P_{n}^{( c, d)}(z) = 2^{-n}  \sum_{p = 0}^{n}(-1)^{n - p} 
\left(\matrix {
n + c  \cr
p \cr
} \right) 
\left(\matrix {
n + d  \cr
n - p \cr
} \right) 
(1 - z)^{n - p}(1-z)^{p} 
\label{s45}
\end{equation}
and
\begin{equation}
P_{n}^{( c, d)}(z) = \frac{\Gamma(n + c + 1)}{n! \Gamma(n + c + d + 1)}  \sum_{r  = 0}^{n} 
\left(\matrix {
n  \cr
r \cr
} \right) 
\frac{\Gamma(n + c + d + r + 1)}{r + c + 1}\left(\frac{z - 1}{2} \right)^{r}
\label{s46}
\end{equation}
where $\left(\matrix {
n  \cr
r \cr
} \right) 
= \frac{n!}{r!(n - r)!} = \frac{\Gamma(n + 1)}{\Gamma(r + 1)\Gamma(n - r + 1)} 
$.

We obtained $\left[ P_{n}^{( c, d)}(1 - 2z) \right]^{2}$ from equations (\ref{s45}) and (\ref{s46})  as
\begin{equation}
\left[ P_{n}^{(c, d)}(1 - 2z)\right]^{2} = z^{n + r - p}(1-z)^{p}A_{nq}(p, r)
\label{s47} 
\end{equation}
and putting equation(\ref{s47}) into equation(\ref{s46}), we have
\begin{equation}
\frac{N_{n_{r}}^{2}A_{nq}(p, r)}{- 2i \alpha}\int_{0}^{1}z^{n + r - p + d}(1-z)^{p + c - 1}dz = 1.
\label{s48}
\end{equation}
Using the definition of hypergeometric function
\begin{equation}
\int_{0}^{1} z^{a - 1}(1 - z)^{-b}ds = \frac{1}{a} [~_{2}F_{1}(a, b; a + 1; 1)], 
\label{s49}
\end{equation}
we obtained $N_{n_{r}}$ as
\begin{equation}
N_{n_{r}} = \sqrt{\frac{B_{nq}(p, r)}{A_{nq}(p, r)}}
\label{s50} 
\end{equation}
where
\begin{eqnarray}
&\displaystyle{A_{nq}(p, r) = \frac{(-1)^{n}[\Gamma(n + c + 1)]^{2}\Gamma(n + d + 1)}{\Gamma(n + c - p + 1)\Gamma(n + d + 1)\Gamma(c + d + 1)} \sum_{p = 0}^{n}\frac{(-1)^{p + r} q^{n - p + r} \Gamma(n + c + d + r  + 1)}{p! r! (n - p )!(n - r)! \Gamma(p + d + 1)}}, \nonumber \\
&\displaystyle{B_{nq}(p, r) = \left[ \frac{_{2}F_{1} (\frac{1}{2} [2 + \Lambda] + n - p + r, 1 - i \varepsilon - p; \frac{1}{2}[4 + \Lambda] n - p + r; 1 )}{-i \alpha (\Lambda + [2n - p + r + 1])} \right]^{-1}}, 
\label{s51}
\end{eqnarray}
we have used $c = i \varepsilon, ~\mbox{and}~
d = \frac{\Lambda}{2}$.

\subsection{Pseudopin symmetry solutions of the Dirac equation with the Trigonometry Scarf potential with arbitrary $\kappa$ in D dimension}
In the case of pseudospin symmetry, that is,  the sum  $\sum(r) = V(r) + S(r) = C_{ps}$, we have,
\begin{equation}
\sum(r) =  \frac{(V_{o} - S_{o})}{\sin^{2} \alpha r} 
\label{s52}.
\end{equation}

Substituting equations (\ref{s22}) and (\ref{s52}) into equation (\ref{s20}), the following equation is obtained,
\begin{equation}
\left[ \frac{d^{2}}{d r^{2}} - \frac{ \gamma_{p} \alpha^{2}}{\cos^{2}{\alpha r}} - \varepsilon_{p}^{2} \alpha^{2} + \frac{ \eta_{p} \alpha ^{2}}{ \sin^{2} \alpha r} \right] F_{n_{r},k}(r) = 0,
\label{s53}
\end{equation}
where, we have used the following notations: 
\begin{eqnarray}
&\gamma_{p} = k(k - 1), \nonumber \\
&\varepsilon_{p}^{2}\alpha^{2} = \left(M + E_{n_{r}, k} \right) \left(M - E_{n_{r}, k} + C_{ps} \right), \nonumber \\
&\eta_{p} \alpha^{2} = (V_{o} - S_{o}) \left(M - E_{n_{r}, k} + C_{s} \right). 
\label{s54}
\end{eqnarray}

Using the same procedure (the N-U method)used in solving equation (\ref{s25}), the energy equation is obtained as follows:
\begin{equation}
(M + E_{n_{r}, \kappa})(M - E_{n_{r}, \kappa} + C_{ps}) = -\frac{\alpha^{2}}{4}\left[2 n + D - 2 + \frac{1}{\alpha}\sqrt{\alpha^{2} + (4V_{o} -  S_{o})(M - E_{n_{r}} + C_{ps})} \right]^{2}
\label{s55},
\end{equation}
where we have defined a principal quantum number as $n - \ell = 2n_{r} + 1$, chosen $k = l + \frac{(D - 1)}{2}$  and $n = 1,2, ~~\ldots $. The explicit values of the energy for different values of $n_{r}$, $\alpha$ and $D$ are shown for pseudospin symmetry in the Table 2. below:

\begin{table}[!hbp]
\caption{Bound-state energy level $E_{n_{r}}$ for different values of $n_{r}$, $\alpha$ and $D$ with $M = C_{p} = V_{o} = S_{o} = 1$.  \vspace*{13pt}} {\small
\begin{tabular}{|c|c|c|c|c|c|c|}\hline
{}&{}&{} &{} &{}&{} &{}\\[-1.5ex]
$D$&$n_{r}$&$\alpha =  0.0001$ & $\alpha = 0.0005$ & $\alpha = 0.001$ & $\alpha = 0.005$&$\alpha = 0.01$\\[2ex]\hline
  &1          &-2.00000001      &-2.00000033       &-2.00000133      &-2.00003333       &-2.00013332\\[1ex]
  &2          &-2.00000003      &-2.00000075       &-2.00000300      &-2.00007500       &-2.00030000\\[1ex]
3 &3          &-2.00000005      &-2.00000133       &-2.00000533      &-2.00013333       &-2.00053324\\[1ex]
  &4          &-2.00000008      &-2.00000208       &-2.00000833      &-2.00020832       &-2.00083310\\[1ex]
  &5          &-2.00000012      &-2.00000300       &-2.00001200      &-2.00030000       &-2.00119952\\[1ex] \hline
  &1          &-2.00000002      &-2.00000052       &-2.00000208      &-2.00005208       &-2.00020832\\[1ex] 
  &2          &-2.00000004      &-2.00000100       &-2.00000408      &-2.00010207       &-2.00040828\\[1ex]
4 &3          &-2.00000006      &-2.00000168       &-2.00000675      &-2.00016874       &-2.00067484\\[1ex]
  &4          &-2.00000010      &-2.00000252       &-2.00000101      &-2.00025206       &-2.00100799\\[1ex]
  &5          &-2.00000014      &-2.00000352       &-2.00000141      &-2.00035204       &-2.00140767\\[1ex] \hline
  &1          &-2.00000003      &-2.00000075       &-2.00000300      &-2.00007499       &-2.00030000\\[1ex]
  &2          &-2.00000005      &-2.00000133       &-2.00000533      &-2.00013333       &-2.00053324\\[1ex]
5 &3          &-2.00000008      &-2.00000208       &-2.00000833      &-2.00020832       &-2.00083310\\[1ex]
  &4          &-2.00000012      &-2.00000252       &-2.00001200      &-2.00003000       &-2.00119952\\[1ex]
  &5          &-2.00000016      &-2.00000408       &-2.00001633      &-2.00040828       &-2.00163245\\[1ex] \hline
  &1          &-2.00000004      &-2.00000102       &-2.00000408      &-2.00010208       &-2.00040828\\[1ex]
  &2          &-2.00000007      &-2.00000168       &-2.00000674      &-2.00016874       &-2.00067485\\[1ex]
6 &3          &-2.00000010      &-2.00000252       &-2.00001008      &-2.00025206       &-2.00100799\\[1ex]
  &4          &-2.00000014      &-2.00000352       &-2.00001408      &-2.00035204       &-2.00140767\\[1ex]
  &6         & -2.00000019      &-2.00000469       &-2.00001874      &-2.00046868       &-2.00187383\\[1ex] \hline
  \end{tabular}\label{tab2} }
\vspace*{-13pt}
\end{table}

We can consequently obtain the wave function $F_{n_{r},\kappa}$ and $G_{n_{r},\kappa}$, respectively as
\begin{equation}
F_{n_{r},k}(z) = C_{n_{r}}(1 - z)^{i \varepsilon_{p}}z^{\frac{1 + \Lambda_{p}}{4}}P_{n_{r}}^{(i \varepsilon_{p}, \frac{\Lambda_{p}}{2} )}(1 - 2z),
 \label{s56} 
\end{equation}
\begin{eqnarray}
G_{n_{r}, k} (z) =      \left\{
\begin{array}{ll}
& C_{n_{r}}(1 - z)^{\frac{i \varepsilon_{p}}{2}}z^{\frac{1 + \Lambda_{p}}{4}}P_{n_{r}}^{(i \varepsilon_{p}, \frac{\Lambda_{p}}{2} )}( 1 - 2z)\left[ - \alpha \left\{\left(i \frac{\Lambda_{p} + 1}{2} \right)\left( \frac{1 - z}{\sqrt{z}} \right) + \varepsilon_{p} \sqrt{z} - \frac{k}{i \tanh{\alpha }} \right\} \right] \\
& \displaystyle{+ \ C_{n_{r}}(1 - z)^{\frac{i \varepsilon_{p} + 2}{2}} z^{\frac{3 + \Lambda_{p}}{4}}P_{n_{r} - 1}^{\left(i \varepsilon_{p} + 1, \frac{\Lambda_{p}}{2} + 1 \right)}(1 - 2z) \left[- \alpha \left\{ \left(i \frac{\Lambda_{p} + 2n_{r}}{2} \right) + \varepsilon_{p}  \right\} \right] }~~,
\end{array}\right.  
\label{d57}
\end{eqnarray}
where $\Lambda_{p} = \sqrt{1  + \eta_{p}}$.

The normalization constant $C_{n_{r}}$ is obtained as
\begin{equation}
C_{n_{r}} = \sqrt{\frac{D_{nq}(p, r)}{E_{nq}(p, r)}}
\label{s58} 
\end{equation}
where
\begin{eqnarray}
&\displaystyle{D_{nq}(p, r) = \frac{(-1)^{n}[\Gamma(n + c + 1)]^{2}\Gamma(n + d + 1)}{\Gamma(n + c - p + 1)\Gamma(n + d + 1)\Gamma(c + d + 1)} \sum_{p = 0}^{n}\frac{(-1)^{p + r} q^{n - p + r} \Gamma(n + c + d + r  + 1)}{p! r! (n - p )!(n - r)! \Gamma(p + d + 1)}}, \nonumber \\
&\displaystyle{E_{nq}(p, r) = \left[ \frac{_{2}F_{1} (\frac{1}{2} [2 + \Lambda_{p}] + n - p + r, 1 - i \varepsilon_{p} - p; \frac{1}{2}[4 + \Lambda_{p}] n - p + r; 1 )}{-i \alpha (\Lambda_{p} + [2n - p + r + 1])} \right]^{-1}}, 
\label{s59}
\end{eqnarray}
we have used $c = i \varepsilon_{p}, ~\mbox{and}~
d = \frac{\Lambda_{p}}{2}$.

\subsection{Special cases}
\subsubsection{PT Symmetric trigonometric Scarf potential}

We consider a case in which $\alpha$ of equation (\ref{s1}) as a pure imaginary parameter, that is, $\alpha \rightarrow i\alpha$, then the potential becomes \cite{Yes07} 
\begin{equation}
V(r) = \frac{S_{o} - V_{o}}{\sinh^2{\alpha r}}.
\label{s60}
\end{equation}
For the spin and pseudospin symmetry conditions, the corresponding energy eigenvalues equation are obtained, respectively as, 
\begin{equation}
(M - E_{n_{r}, \kappa})(M + E_{n_{r}, \kappa} - C_{s}) = \frac{\alpha^{2}}{4}\left[ 2 n + D - \frac{i}{ \alpha} \sqrt{4(S_{o} + V_{o}) \left(M + E_{n_{r}, \kappa} - C_{s} \right) - \alpha^{2}} \right]^{2}
\label{s61}
\end{equation}
and
\begin{equation}
(M + E_{n_{r}, \kappa})(M - E_{n_{r}, \kappa} + C_{ps}) = \frac{\alpha^{2}}{4}\left[ 2 n + D - 2 - \frac{i}{ \alpha} \sqrt{4(S_{o} - V_{o}) \left(M - E_{n_{r}, \kappa} + C_{ps} \right) - \alpha^{2}} \right]^{2}.
\label{s62}
\end{equation}

\subsubsection{PT Symmetric and $q$-deformed trigonometric Scarf potential}

In equation (\ref{s1}), we have used a mapping $r\rightarrow r -\frac{1}{\alpha} \ln{ \sqrt{q}}$\cite{Yes07, deS05}, then, the potential becomes 
\begin{equation}
V_{q}(r) = \frac{S_{0} - V_{0}}{\sinh_{q}^2 \alpha r}
\label{s63}
\end{equation}
and the energy equations for spin and pseudospin symmetry of the $q$-deformed trigonometric Scarf potential are obtained, respectively as:
\begin{equation}
(M - E_{n_{r}, \kappa})(M + E_{n_{r}, \kappa} - C_{s}) = \frac{\alpha^{2}}{4}\left[ 2 n + D  - \frac{i}{\alpha} \sqrt{4q^{-1}(S_{o} + V_{o})(M + E_{n_{r}, \kappa} - C_{s})} - \alpha^{2}\right]^{2}.
\label{s64}
\end{equation}
and
\begin{equation}
(M + E_{n_{r}, \kappa})(M - E_{n_{r}, \kappa} + C_{ps}) = \frac{\alpha^{2}}{4}\left[ 2 n + D - 2 - \frac{i}{\alpha} \sqrt{4q^{-1}(S_{o} - V_{o})(M - E_{n_{r}, \kappa} + C_{ps})} - \alpha^{2}\right]^{2}
\label{s65}
\end{equation}

\subsubsection{ Non-PT Symmetric, non Hermitian and $q$-deformed trigonometric Scarf potential}

Another form of the potential in equation (\ref{s1}) is obtained by putting $S_{0} = S_{1} + i S_{2}$ and $V_{0} = V_{1} + i V_{2}$, $\alpha = i \alpha$ and $q  \rightarrow i q$ \cite{Yes07}. In this case, after some manipulations, the potential takes the form
\begin{equation}
V_{q}(r) = 4 \frac{(S_{1} - V_{1}) + i (S_{2} - V_{2})}{(e^{r} - i q e^{-r})^2}.
\label{s66}
\end{equation}
Then, the energy equations of the spin and pseudospin symmetry for this potential are obtained, respectively as:
\begin{eqnarray}
&\displaystyle{(M - E_{n_{r}, \kappa})(M + E_{n_{r}, \kappa} - C_{s})} \nonumber \\
&\displaystyle{ = \frac{\alpha^{2}}{4} \left[ 2 n + D - \frac{i}{\alpha} \sqrt{4q^{-1} [(V_{1} + S_{1}) i + (S_{2} + V_{2})](M + E_{n_{r}, \kappa} - C_{s}) - \alpha^{2}} \right]^{2}}
\label{s67}
\end{eqnarray}
and
\begin{eqnarray}
&\displaystyle{(M + E_{n_{r}, \kappa})(M - E_{n_{r}, \kappa} + C_{ps})} \nonumber \\
&\displaystyle{ = \frac{\alpha^{2}}{4} \left[ 2 n + D - 2 - \frac{i}{\alpha} \sqrt{4q^{-1} [(V_{1} - S_{1}) i + (S_{2} - V_{2})](M - E_{n_{r}, \kappa} + C_{ps}) - \alpha^{2}} \right]^{2}}
\label{s68}.
\end{eqnarray}

\section{Conclusion}

In this work, we have obtained the bound state solutions of the $\textit{D}$-dimensional Dirac equation with spin and pseudospin symmetry for scalar and vector trigonometric Scarf potential. The two -component spinors and corresponding the energy equation have been obtained using Nikiforov-Uvarov method. The numerical results shows that the bound state energy level $E_{n_{r},\kappa}$ for spin and pseudospin symmetry increases with both dimensions $D$ and range parameter $\alpha$. In addition, PT symmetric, $q$-deformed, non-PT symmetric and non-Hermitian version of this potential are investigated in this study as special cases and their energy equations are obtained.

\end{document}